\documentclass[12pt,superscriptaddress,aps,prd,preprint]{revtex4}
\usepackage{amsmath}
\usepackage{amssymb}
\makeatletter
\newcommand{\bea}{\begin{eqnarray}}
\newcommand{\eea}{\end{eqnarray}}

\newcommand{\pa}{\partial}

\usepackage[dvips]{graphicx}

\begin{document}
\title{Dynamical Chern-Simons modified gravity, G\"{o}del Universe and variable cosmological constant}
\author{C. Furtado}
\affiliation{Departamento de F\'{\i}sica, Universidade Federal da Para\'{\i}ba\\
Caixa Postal 5008, 58051-970, Jo\~ao Pessoa, Para\'{\i}ba, Brazil}
\email{furtado,jroberto,petrov@fisica.ufpb.br}
\author{J. R. Nascimento}
\affiliation{Departamento de F\'{\i}sica, Universidade Federal da Para\'{\i}ba\\
Caixa Postal 5008, 58051-970, Jo\~ao Pessoa, Para\'{\i}ba, Brazil}
\email{furtado,jroberto,petrov@fisica.ufpb.br}
\author{A. Yu. Petrov}
\affiliation{Departamento de F\'{\i}sica, Universidade Federal da Para\'{\i}ba\\
Caixa Postal 5008, 58051-970, Jo\~ao Pessoa, Para\'{\i}ba, Brazil}
\email{furtado,jroberto,petrov@fisica.ufpb.br}
\author{A. F. Santos}
\affiliation{Instituto Federal de Educa\c {c}\~{a}o, Ci\^{e}ncia e Tecnologia de Mato Grosso, Campus C\'{a}ceres,
Av. dos Ramires, S/N, Distrito Industrial, 78200-000,
C\'{a}ceres, MT, Brazil}
\email{alesandrodossantos@gmail.com}

\begin{abstract}
We study the condition for the consistency of the G\"{o}del metric with the dynamical Chern-Simons modified gravity. It turns out to be that this compatibility can be achieved only if the cosmological constant is variable in the space.
\end{abstract}

\maketitle

The problem of breaking of the Lorentz and CPT symmetries is intensively studied now. Among a great number of possible Lorentz-CPT breaking modification of gravity \cite{Kost}, the four-dimensional gravitational Chern-Simons term originally introduced in \cite{JaPi} is certainly the most popular one. The main reasons for it are its relatively simple form, evident gauge symmetry and straightforward analogies with the three-dimensional gravitational Chern-Simons term \cite{DJT}. A lot of issues treating different aspects of the four-dimensional gravitational Chern-Simons term has been studied, in particular, in \cite{ours} it was shown to emerge as a quantum correction. The most important line of studies of this term is actually devoted to searching for the solutions of the equations of motion in the modified theory whose action is a sum of the usual Einstein-Hilbert action and of the gravitational Chern-Simons term. It was shown that a lot of well-known solutions of the usual Einstein equations, in particular, spherically symmetric and cylindrically symmetric ones, such as Schwarzschild, Reissner-Nordstrom and Friedmann-Robertson-Walker metrics resolve the modified Einstein equations as well \cite{Grumiller}. Further, in \cite{Go} the compatibility of the G\"{o}del metric \cite{Godel} with modified Einstein equation also has been shown (for some other issues related to the G\"{o}del metric, see also \cite{Reb}).

However, the Kerr metric failed to solve the modified Einstein equations. To circumvent this difficulty, in \cite{Grumiller,Erick}, the concept of the dynamical Chern-Simons coefficient was introduced. Within this formulation, where the function $\theta$ multiplying the topological ${}^*RR$ term in the modified gravity action is considered as a dynamical variable, it was shown that the Kerr metric must be adequately modified to satisfy the new modified equations of motion involving dynamics of $\theta$ as well \cite{Konno}. Therefore, a natural problem emerges -- whether other known metrics, or, probably, other parameters of the theory must suffer any modification to solve the modified equations of motion? In this paper we are going to consider this problem for the G\"{o}del metric, and our result turns out to be highly nontrivial -- indeed, we find that this metric solves new modified equations of motion only in the case of variable cosmological constant! This effect can be treated as a justification of the hypotheses of the variable cosmological constant which are very popular in the context of search for a possible explanation of accelerated expansion of the Universe (see f.e. \cite{Sola}).

The starting point of our consideration is the following action for the Chern-Simons modified gravity:
\bea
\label{smod}
S=\frac{1}{16\pi G}\int d^4x\Bigl[\sqrt{-g}(R-\Lambda)+\frac{l}{64\pi}\theta \,{^*}RR-\frac{1}{2}\pa^\mu\theta\pa_\mu\theta\Bigl] + S_{mat},
\eea
where, unlike \cite{Go}, the function $\theta$ is a dynamical variable \cite{Erick}.  
Varying this action with respect to the metric and to the scalar field $\theta$, we obtain the following equations of motion:
\bea
&& G_{\mu\nu}+lC_{\mu\nu}= T_{\mu\nu}, \\
&&g^{\mu\nu}\nabla_{\mu}\nabla_{\nu}\theta=-\frac{l}{64\pi} {^*}RR,\label{2}
 \label{einst}
\eea
where $G_{\mu\nu}$ is the Einstein tensor (within this formulation, the contribution from the cosmological term is absorbed into the energy-momentum tensor) and $C_{\mu\nu}$ is the Cotton tensor defined as
\bea
C^{\mu\nu}=-\frac{1}{2\sqrt{-g}}\Bigl[v_\sigma \epsilon^{\sigma\mu\alpha\beta}D_\alpha R^\nu_\beta+\frac{1}{2}v_{\sigma\tau}\epsilon^{\sigma\nu\alpha\beta}R^{\tau\mu}\,_{\alpha\beta}\Bigl] \,+\, (\mu\longleftrightarrow\nu), 
\eea
and ${^*}RR$ is the topological invariant called the Pontryagin term whose explicit form is
\bea
{^*}RR\equiv {^*}{R^a}\,_b\,^{cd}R^b\,_{acd},
\eea
where $R^b\,_{acd}$ is the Riemann tensor and ${^*}{R^a}\,_b\,^{cd}$ is the dual Riemann tensor given by
\bea
{^*}{R^a}\,_b\,^{cd}=\frac{1}{2}\epsilon^{cdef}R^a\,_{bef}.
\eea
In this paper we are concentrated on the case of the G\"{o}del metric which is written as \cite{Godel}
\bea
ds^2=a^2\Bigl[dt^2-dx^2+\frac{1}{2}e^{2x}dy^2-dz^2+2 e^x dt\,dy\Bigl],\label{godel}
\eea
where $a$ is a positive number.

The energy-momentum tensor $T_{\mu\nu}$ in this theory is composed of two terms: 
\bea
T_{\mu\nu}=T_{\mu\nu}^m+T_{\mu\nu}^{\theta}, 
\eea
where $T_{\mu\nu}^m$ is the energy-momentum tensor of an usual matter (including the contribution from the cosmological term). We suggest that this energy-momentum tensor is the same one that yields G\"{o}del solution in the simple Einstein gravity \cite{Godel} as well as in the Chern-Simons modified gravity with the external $\theta$ parameter \cite{Go}, that is,
$T_{\mu\nu}^m=8\pi\rho u_\mu u_\nu+\Lambda g_{\mu\nu}$,  where $u$ is a unit time-like vector whose explicit contravariant components look like $u^{\mu}=(\frac{1}{a},0,0,0)$  and the corresponding covariant components are $u_\mu=(a,0,ae^x,0)$. Further, the energy-momentum tensor for the $\theta$ field looks like
\bea
T_{\mu\nu}^{\theta}=(\partial_{\mu}\theta)(\partial_{\nu}\theta)-\frac{1}{2}g_{\mu\nu}(\partial^{\lambda}\theta)(\partial_{\lambda}\theta),
\eea
that is, it reproduces the form of the usual energy-momentum tensor for the scalar field in a curved space. 

We have showed in \cite{Go}, that the choice $\theta=\theta(x,y)$ annihilates the Cotton tensor. Let us verify a compatibility of the G\"{o}del metric with the equations of motion (\ref{einst}) just in this case simplifying all expressions drastically. 
For the $\theta$ dependent only on $x$ and $y$, the non-zero components of the $T_{\mu\nu}^{\theta}$ look like
\bea
T_{00}^\theta&=&\frac{1}{2}(\partial_1\theta)^2+e^{-2x}(\partial_2\theta)^2\nonumber\\
T_{02}^\theta&=&\frac{1}{2}e^x(\partial_1\theta)^2+e^{-x}(\partial_2\theta)^2\nonumber\\
T_{11}^\theta&=&\frac{1}{2}(\partial_1\theta)^2-e^{-2x}(\partial_2\theta)^2\nonumber\\
T_{12}^\theta&=&(\partial_1\theta)(\partial_2\theta)\nonumber\\
T_{22}^\theta&=&\frac{1}{4}e^{2x}(\partial_1\theta)^2+\frac{1}{2}(\partial_2\theta)^2\\
T_{33}^\theta&=&-\frac{1}{2}(\partial_1\theta)^2-e^{-2x}(\partial_2\theta)^2.
\eea

Taking into account the expression for the Einstein tensor from \cite{Godel}, that is,
\bea
G_{00}=\frac{1}{2}, \, G_{11}=\frac{1}{2}, \, G_{20}=\frac{1}{2}e^x, \, G_{22}=\frac{3}{4}e^{2x}, \, G_{33}=\frac{1}{2}, 
\eea
one can write down the nontrivial components of the Einstein equations, 00, 11, 22 and 33 respectively:
\bea
\label{comp}
\frac{1}{2}&=&8\pi\rho a^2+\Lambda a^2+\left[\frac{1}{2}(\partial_1\theta)^2+e^{-2x}(\partial_2\theta)^2\right],\label{00}\\
\frac{1}{2}&=&-\Lambda a^2+\left[\frac{1}{2}(\partial_1\theta)^2-e^{-2x}(\partial_2\theta)^2\right],\label{11}\\
\frac{3}{2}&=&16\pi\rho a^2+\Lambda a^2+\left[\frac{1}{2}(\partial_1\theta)^2+e^{-2x}(\partial_2\theta)^2\right],\label{22}\\
\frac{1}{2}&=&-\Lambda a^2-\left[\frac{1}{2}(\partial_1\theta)^2+e^{-2x}(\partial_2\theta)^2\right].\label{33}
\eea
The equations for the components 00 and 02 turn out to coincide identically.  Beside of this, the component 12 of modified Einstein equations is
\bea
(\partial_1\theta)(\partial_2\theta)=0.\label{12}
\eea

A straightforward inspection of these components allows to conclude that solutions of the equations for the components $00,02,22,33$ can be formally written as
\bea
8\pi\rho=\frac{1}{a^2};\,\,\,\,\,\,\,\,\,\,\,\,\,\,\,\,\,\, \,\,\,\,\,\,\,\,\,\,\,\,\,\,\,\,\,\,\Lambda=-\frac{1}{2a^2}-\frac{\Theta}{a^2},\label{17}
\eea
where
\bea
\Theta=\frac{1}{2}\left(\partial_1\theta\right)^2+e^{-2x}\left(\partial_2\theta\right)^2.
\eea
Therefore we conclude that these solutions correspond to the case when the cosmological constant is not a constant more but an external field dependent on the space-time coordinates which can be considered as a constant only approximately. However, this does not modify derivation of the corresponding term of the Einstein equations since the $\Lambda$ (and, as a consequence, $\Theta$) does not depend on the metric tensor, hence the formal structure of the Einstein equations is the same with only difference is in the fact that the cosmological "constant" $\Lambda$ is now not a constant but a fixed function which does not depend on the metric tensor (the possibility of the spatial dependence of $\Lambda$ has been argued in a similar way in \cite{Massa}). 

It is easy to see that in the limit $\Theta\to 0$ we recover the usual solution for the G\"{o}del universe \cite{Godel}. However, to proceed further, one must take into account that the condition (\ref{17}) does not solve the equation (\ref{11}) for the component 11. Returning again to the equation (\ref{12}) we find that one must choose either $(\partial_1\theta)=0$ or $(\partial_2\theta)=0$.
If, for example, we choose $(\partial_1\theta)=0$, we conclude that $\theta=\theta(y)$. Thus, the $\Theta$ is reduced as  
\bea
\Theta\rightarrow\bar{\Theta}=e^{-2x}(\partial_2\theta)^2.
\eea
We note that the choice $(\partial_2\theta)=0$ implies only in the trivial case $\theta=const$.

As a result, we conclude that the G\"{o}del metric can be solution of the equations of motion in the Chern-Simons modified gravity with dynamical $\theta$ field if the cosmological constant is really a slowly varying field of the form
\bea
\Lambda=-\frac{1}{2a^2}(1+2\bar{\Theta}).\label{Lambda}
\eea

To achieve the complete consistency of the equations of motion, we must solve also the equation (\ref{2}) which allows us to find the explicit form of the scalar field. Since for the G\"{o}del metric ${^*}RR=0$ (indeed, ${^*}RR=\frac{1}{2}\epsilon^{abcd}R_{ab}^{\phantom{ab}mn}R_{cdmn}$; at the same time, at least one of the indices $a,b,c,d$ must be equal to 3, but all components of the Riemann tensor with at least one index equal to 3 evidently vanish, $R_{3bcd}=0$), this equation is reduced to
\bea
g^{\mu\nu}\nabla_\mu\nabla_\nu\theta=0.
\eea
Taking into account that, as we have concluded, the $\theta$ depends only on $y$ coordinate, we have 
\bea
\partial_y^2\theta(y)=0,
\eea
which yields
\bea
\theta(y)=c_1+c_2y,
\eea
where $c_1$ and $c_2$ are some constants.

Substituting this expression for $\theta$ into (\ref{Lambda}), we obtain the following result for $\Lambda$:
\bea
\Lambda=-\frac{1}{2a^2}\left[1+C\,e^{-2x}\right],
\eea
where $C$ is a constant. Therefore, the cosmological "constant" is a function of the spatial coordinate, i.e., $\Lambda=\Lambda(x)$. One should note that, in principle, the spatial dependence of the  cosmological constant does not create any conceptual difficulties from the viewpoint of the Lagrangian formalism. Indeed, had we considered the action
\bea
S=\frac{1}{16\pi G}\int d^4x\Bigl[\sqrt{-g}(R-\Lambda(x))+\frac{l}{64\pi}\theta \,{^*}RR-\frac{1}{2}\pa^\mu\theta\pa_\mu\theta\Bigl] + S_{mat},
\eea
with $\Lambda(x)$ is a given function of the space-time coordinates, with no dependence on the metric tensor, we would obtain the same equations of motion (\ref{einst}) as in the usual case of the constant $\Lambda$.

Let us discuss our results. The key conclusion we obtain, with no doubts, is a fact that the modified Einstein gravity with a dynamical Chern-Simons coefficient admits the G\"{o}del solutions, as well as the theory where the Chern-Simons coefficient is treated as an external field \cite{Go}. However, we find that, first, the structure of the Chern-Simons coefficient compatible with the G\"{o}del metric is strongly restricted, that it, it is at most linear in the coordinate $y$, second, that, to satisfy the modified Einstein equations, the cosmological constant cannot be constant in the strong sense of this word, it depends on the same coordinate $x$ on which the components of the G\"{o}del metric depend. One must remind that the idea of variability of the cosmological constant, which is treated now as one of the possible explanation of the accelerated expansion of the Universe \cite{Lima} has a long story, but most part of applications of this idea were based on the suggestion that the $\Lambda$ depends only on time, which is reasonable within the cosmological framework. Nevertheless, the spatial dependence of the cosmological constant could imply in highly nontrivial modifications of the topology of the space, so, the space turns out to be closed on itself around a massive object \cite{Narlikar} (for other issues related to the spatial dependence of $\Lambda$, see also \cite{Massa} and references therein; it should be mentioned that in \cite{Massa} it was claimed that the spatial dependence of the cosmological constant can be treated as a possible solution of the Pioneer anomaly problem). The natural problem following from this fact is whether in this case the closed spacetime curves whose presence is characteristic for the G\"{o}del metric \cite{Godel} can arise. Other possible consequences of the spatial dependence of a cosmological constant are the deep relations between electromagnetic and gravitational interactions implying in so-called electromagnetic mass models \cite{Ray} and in possibility for existence of new solutions with nontrivial topology, such as cosmic strings and wormholes (see f.e. \cite{Put}). As a result, we can conclude that the consistency of the model of the modified gravity with a dynamical Chern-Simons term and the consequent variability of $\Lambda$ open new horizons for studies in gravity.

{\bf Acknowledgments.}
C. F., J. R. N. and A. Yu. P. have been supported by Conselho Nacional de
Desenvolvimento Cient\'\i fico e Tecnol\'ogico (CNPq) and Coordena\c c\~ao de Aperfei\c coamento de Pessoal de N\'\i vel Superior (CAPES: AUX-PE-PROCAD 579/2008). A. Yu. P. has been supported by the CNPq project No. 303461-2009/8.

\end{document}